\documentstyle[aps,preprint,epsfig]{revtex} 
\begin{document}
\draft                   
\title{Calculation of thermodynamic properties of finite Bose-Einstein systems}
\author{Peter Borrmann, Jens Harting, 
Oliver M{\"u}lken, and Eberhard R.~Hilf}
\address{Department of Physics, Carl von Ossietzky University Oldenburg,
D-26111 Oldenburg, Germany}
\date{\today}
\maketitle
\begin{abstract}
We derive an exact recursion formula for the calculation of
thermodynamic functions of finite systems obeying Bose-Einstein
statistics. The formula is applicable for canonical systems where the
particles can be treated as noninteracting in some approximation, e.g.
like Bose-Einstein condensates in magnetic traps. The numerical
effort of our computation scheme grows only linear with the 
number of particles.\\
As an example we calculate the relative ground state fluctuations
and specific heats for ideal Bose gases with a finite numbers of
particles enclosed in containers of different shapes. 
\end{abstract}
\pacs{03.75.Fi, 05.30.Jp, 32.80.Pj}
With the observation of Bose-Einstein condensation (BEC) of
magnetically \cite{Hulet95,Cornell95,Ketterle95} and optically 
\cite{Ketterle98} trapped atoms  
new insights into the nature of this state of matter have been 
given. The experimental situation is in all cases quite different 
from the ideal gas treated within the grand canonical ensemble, 
which is the standard textbook example. First, the number of
particles within the traps is fixed and finite, which suggests a 
canonical or microcanonical treatment of the systems. Second, the 
confining trap potentials greatly influence the condensate
properties. Third, although the trapped gases are quite dilute 
the validity of the treatment as non-interacting particle gases has 
to be checked from case to case.\\
Even within the approximation of non-interacting particles the 
calculation of the thermodynamic properties of the Bose-Einstein 
systems remains a hard to tackle mathematical problem. 
Recently, some approximate methods to calculate the fluctuation of
the ground state occupation number in a trapped Bose-Einstein
condensate have been developed
\cite{Holthaus97,Holthaus97a,Rzazewski97,Rzazewski97a}.  Here we present an
exact method to calculate all thermodynamic quantities of finite
canonical Bose systems, given the one particle density of states.

As the starting point we utilize the recursive formula of the canonical
partition function for a system of $N$ noninteracting bosons as given in
\cite{bf93}:
\begin{equation} \label{recur_can}
Z_N(\beta) = \frac{1}{N} \sum_{k=1}^{N} Q_k(\beta) Z_{N-k}(\beta) \;,
\end{equation}
where $Q_k(\beta) = Z_1(k \beta)= \sum_{i} \exp(-k \beta \epsilon_i)$ is
the one-particle partition function at the temperatures $k\beta$ and
$Z_0(\beta)=1$. The microcanonical partition $\Gamma_N(E)$ can be
calculated by an inverse Laplace-transform of (1) and is given by 
\begin{eqnarray}
\Gamma_N(E)& =& \frac{1}{N} \sum_{k=1}^{N} 
\frac{1}{2 \pi i} \int_{c-i \infty}^{c+i \infty}
{\rm d}\beta \exp(\beta E) Q_k(\beta) Z_{N-k}(\beta) \\
&=& \frac{1}{N} \sum_{k=1}^{N}  \nonumber
\int_{0}^{E} {\rm d}E' \Gamma_1^k(E') \Gamma_{N-k}(E-E') \;,
\end{eqnarray}
where $\Gamma_1^k(E)$ is the inverse Laplace-transform of $Q_k(\beta)$
and $\Gamma_0(E) = \delta(E)$. A similar, slightly less
general, equation has recently been derived by Weiss and Wilkens
\cite{Wilkens97a}.

Eq.(\ref{recur_can}) can be used to calculate all thermodynamic
quantities by appropriate differentiation of $\ln Z_N$. However, in any
case $Z_N$ occurs as a normalization factor and has to be calculated
explicitly. This turns out to be a major drawback. First, the numerical
effort to calculate $Z_N$ grows with the square of the particle number
$N$. 
Moreover, since $Z_N(\beta)$ grows exponentially with $N$ multiple
precision arithmetic is required for proper calculation. 
We will present a method avoiding these difficulties. 

To ease our derivations we rewrite $Z_N(\beta)$ utilizing the
$\cal{Z}$-transform and define:
\begin{eqnarray}
{\cal Z}(Z)&=& F(x) = \sum_{k=0}^{\infty} \frac{Z_k(\beta)}{x^{k}},\\
{\cal Z}(Q)&=& G(x) = \sum_{k=0}^{\infty} \frac{Q_k(\beta)}{x^{k}},
\end{eqnarray}
where we define $Q_0(\beta) = 0$.  Taking advantage of the basic
properties of the $\cal{Z}$-transform eq.(\ref{recur_can}) can be
written in the form
\begin{equation}
-x \frac{\rm{d}}{\rm{d}x} F(x) = F(x) G(x),
\end{equation}
yielding \footnote{Note that $F(x)$ is closely related to the grand
canonical partition function.}
\begin{equation} 
F(x)=\exp\left(\sum_{k=1}^{\infty} \frac{Q_k(\beta)}{k} x^{-k} \right).
\end{equation}
Applying the inverse $\cal{Z}$-transform we may write $Z_N(\beta)$ as 
\begin{equation}
Z_N(\beta) = \frac{1}{2 \pi i} \int_C F(x) x^{N-1} {\rm d}x,
\end{equation}
where $C:=\{z\in C: \mid x \mid = r\}$ and $r$ has to satisfy the 
condition $\mid Z_N(\beta)\mid \le \exp(r N) $.
Alternatively we may write:
\begin{equation} \label{diff_z}
Z_N(\beta) = \frac{1}{N!} \left. \frac{\rm{d}^N}{\rm{d}x^N} 
F(1/x)\right|_{x=0}.
\end{equation}
Using eq.(\ref{diff_z}) the number of particles  with energy
$\epsilon_i$ can be calculated by
\begin{eqnarray}
\eta_i(N,\beta)&=&\frac{-1}{\beta}
\frac{\partial}{\partial \epsilon_i} 
\ln Z_N(\beta)\\
&=&  \frac{1}{Z_N(\beta)} \sum_{k=1}^{N} 
\exp(-\beta k \epsilon_i) Z_{N-k}(\beta) \nonumber
\end{eqnarray}
Some reordering yields:
\begin{equation} \label{recocc}
\eta_i(N+1,\beta) = \frac{Z_N(\beta)}{Z_{N+1}(\beta)} 
\exp(-\beta \epsilon_i) (\eta_i(N,\beta)+1) \; .
\end{equation}
Since the particle number is a conserved quantity in the canonical
ensemble the direct calculation of the normalization factor 
can be omitted by using the relation
\begin{equation} \label{norm}
\frac{Z_N(\beta)}{Z_{N+1}(\beta)} = 
\frac{N+1}{\sum_{i=0}^{\infty} \exp(-\beta \epsilon_i)
(\eta_i(N,\beta)+1)} \;.
\end{equation}
For Fermi systems the following recursion formula
\begin{equation} \label{recocc_fermi}
\eta_i(N+1,\beta) = \frac{Z_N(\beta)}{Z_{N+1}(\beta)} 
\exp(-\beta \epsilon_i) (1-\eta_i(N,\beta)) \;,
\end{equation}
with
\begin{equation} \label{norm_fermi}
\frac{Z_N(\beta)}{Z_{N+1}(\beta)} = 
\frac{N+1}{\sum_{i=0}^{\infty} \exp(-\beta \epsilon_i)
(1-\eta_i(N,\beta))} \;.
\end{equation}
can be derived in a similar manner.
In practice, only a limited number of energy levels has to be taken
into account, since the occupation probability rapidly decreases
with increasing energy eigenvalues.   Equations (\ref{recocc}) and
(\ref{norm}) are extremely useful in practical calculations. The
numerical effort to calculate the occupation numbers grows only
linear with the number of particles.   Moreover, only a moderate
arithmetic precision is required.  Having the occupation
probabilities at hand the energy expectation value is given by:
\begin{equation}
  E(N,\beta) = \sum_{i=0}^{\infty} \epsilon_i \eta_i(N,\beta) \;.
\end{equation}
The calculation of the fluctuation of the occupation
probabilities $\delta \eta_i(N,\beta)$ is a little bit more
complicated and contains another recursion:
\begin{eqnarray} \label{delta}
&&\left( \delta\eta_i(N+1,\beta) \right)^2 = \frac{1}{\beta^2} 
    \frac{\partial^2}{\partial \epsilon_i^2} \ln(Z_{N+1}(\beta))\\
   &=& \frac{-1}{\beta} \frac{\partial}{\partial \epsilon_i}
       \eta_{i}(N+1,\beta) \nonumber \\
&=& \frac{Z_{N}(\beta)}{Z_{N+1}(\beta)}\exp(-\beta \epsilon_i)
\left\{ \delta^2\eta_i(N,\beta) \right. \nonumber \\
&+& \left. \left[\eta_i(N+1,\beta)+1\right] 
\left[\eta_i(N,\beta)-\eta_i(N+1,\beta)+1 \right]
\right\} \;. \nonumber
\end{eqnarray}

To illustrate the usefulness of our recursion formulas 
we consider the ideal gas with parameters of liquid 
Helium in containers of different shapes :\\
i) a cube with side-length $L_x, L_y, L_z$, and energy levels
\begin{equation}
E_{n_x,n_y,n_z} = \frac{\pi^2 \hbar^2}{2 m_{He}} 
\left( 
\frac{n_{x}^{2}}{L_x^2} +
\frac{n_{y}^{2}}{L_y^2} +
\frac{n_{z}^{2}}{L_z^2} 
\right)\,
\end{equation}
ii) a sphere with radius $a$ and energy levels
\begin{equation}
E_{n,l} = \frac{\hbar^2}{2 m_{He} a^2} u_{n,l}
\end{equation}
and degeneracy $\sigma_{n,l} = 2l+1$, and\\
iii) a cylinder with diameter $d=2a$, height $L$, energy levels
\begin{equation}
E_{n,l,m} =  \frac{\hbar^2}{2 m_{He}}
\left( \frac {v_{n,l}^2}{a^2} + \frac{m^2 \pi^2}{L^2} \right)
\end{equation}
with $n=1,2,\ldots$, $l=1,2,3,\ldots$, and $m=\ldots,-1,0,-1,\ldots$.\\
We denoted the zeros of the half integer Bessel functions
$J_{n+1/2}(r)$ by $u_{n,l}$ and the zeros of the integer
Bessel function $J_{n}(r)$ by $v_{n,l}$.

Fig.~1 displays the specific heats and the fluctuations of the
ground state occupation number $\delta \eta_0/N$ as a function of
the canonical temperature for different trap geometries and
$N=100,1000$, and $10000$ He-atoms. In all cases the particle
density is taken to be $\rho =0.0216 \AA^{-3}$. With growing system
size the differences between the specific heats for the different
trap geometries almost vanish and approach the typical shape of the
curve for the ideal Bose gas. I.e., with respect to the specific
heat the boundary conditions get more and more unimportant with
increasing volume.  In contrast, the ground state fluctuations
exhibit a completely different behaviour. The cubic box, the {\sl
compact} cylinder with equal diameter and height, and the sphere
show almost equal ground state fluctuations for all system sizes,
while the ground state fluctuations of the stretched box and the
stretched cylinder are remarkably larger for temperatures below the
critical temperature. 
This effect is not unexpected because restricting the particle motion
in one or two dimensions makes the system act like a lower dimensional 
system, which are known to  show larger fluctuations.
Moreover, the differences between the
fluctuations of the {\sl stretched} traps and the {\sl compact} traps
do not decrease with increasing system size.  The reason for this
behaviour is found in the energy difference between the ground state
and the first excited level, which is much
larger for the {\sl stretched} traps than for the {\sl compact} 
traps. Since $\delta \eta_0/N$ decreases approximately with
$N^{-1/3}$ the infinite particle number limit is the same for
all trap geometries. Under experimental considerations our results
imply that the stability of the condensate fraction in anisotropic
traps should be considerably smaller than in isotropic traps. In
Fig.~1 we plotted $\delta \eta_0/N$ to allow good comparison with
previous published results \cite{Rzazewski97a,Holthaus97}. Since 
this quantity goes to zero as the system size increases it is a bad
indicator for phase transitions. The relative ground state
fluctuation $\delta \eta_0/\eta_0$ shown in Fig.~2 is much more
conclusive in this respect. 

True phase transition only occur for infinite systems. 
However, it is well known from other
systems, e.g. finite spin lattices and  clusters 
\cite{kunz93,wales94}, that finite systems 
already display the onset of phase transitions. 
Instead of having a well defined critical temperature 
the transition occurs in a broader crossover region.
As can been extracted from 
Fig.~2 the crossover region, which is indicated by the sharp increase
of $\delta \eta_0/\eta_0$, extends even for the cubic box with 
10000 atoms over a temperature range of 0.5 K. For the {\sl
stretched} trap the crossover region is about twice as large.

All calculations have been performed on an IBM-43P (233 MHz)
workstation. For 1000 particles a run with 200 temperature points
took about six minutes, for 10000 particles one hour. A calculation
of the same quantities with the recursion formula given in equation
(\ref{recur_can}) takes for 1000 particles about six hours (and
would take at least 600 hours for 10000 particles).  Moreover, due
to the numerical instabilities connected with (\ref{recur_can})
$N=1000$ was the largest particle number we achieved with this
formula, even though we utilized a multiple precision package.

We expect the recursion formulas to be quite useful for calculating
the properties of dilute atomic Bose gases in magnetic traps with
different geometry. For this propose we provide an easy to use Java
program\footnote{The code is available within the World-Wide-Web:\\
http:$\backslash\backslash$www.physik.uni-oldenburg.de$\backslash\sim
$borrmann$\backslash$BEC}, which requires as input only the energy
level distribution with appropriate degeneracy and calculates the
basic properties of finite Bose systems with particle numbers up to
10000. A slightly faster Fortran code is available upon Email
request\footnote{Corresponding author: {\sl
borrmann@uni-oldenburg.de}}.

\begin{figure}[h]
\caption{(a-c) Specific heats $C_v/N$ and (d-f)  ground state
fluctuations as a function of the canonical temperature for
systems of $N=100$, $N=1000$, and $N=10000$ particles. The solid lines
represent the results for a spherical trap, the dashed lines for a
cube, the circles for a cylinder with a diameter to height ratio of
$d/L=1$, the long dashed lines for a box with side-lengths $L_z=4
L_y=4 L_x$, and the squares for a cylinder with  $d/L=1/4$.  In all
cases the particle density is taken to be $\rho =0.0216 \AA^{-3}$.}
\end{figure}
\begin{figure}[h]
\caption{Relative ground state fluctuations $\delta\eta_0/\eta_0$ 
as a function of temperature for the cubic box 
and the stretched box and $N=1000$, and 10000 particles.} 
\end{figure}


\begin{thebibliography}{10}

\bibitem{Hulet95}
C. Bradley, C. Sackett, J. Tollet, and R. Hulet, Phys. Rev. Lett. {\bf 75},
  1687  (1995).

\bibitem{Cornell95}
M. Anderson {\it et~al.}, Science {\bf 69},  198  (1995).

\bibitem{Ketterle95}
K. Davies {\it et~al.}, Phys. Rev. Lett. {\bf 75},  3969  (1995).

\bibitem{Ketterle98}
D. Stamer-Kurn {\it et~al.}, Phys. Rev. Lett. {\bf 80},  2027  (1998).

\bibitem{Holthaus97}
S. Grossmann and M. Holthaus, Phys. Rev. Lett. {\bf 79},  3557  (1997).

\bibitem{Holthaus97a}
S. Grossmann and M. Holthaus, OPTICS EXPRESS {\bf 1},  262  (1997).

\bibitem{Rzazewski97}
P. Navez {\it et~al.}, Phys. Rev. Lett. {\bf 79},  1789  (1997).

\bibitem{Rzazewski97a}
M. Gajda and K. Rzazewski, Phys. Rev. Lett. {\bf 78},  2686  (1997).

\bibitem{bf93}
P. Borrmann and G. Franke, J. Chem. Phys {\bf 98},  2484  (1993).

\bibitem{Wilkens97a}
C. Weiss and M. Wilkens, OPTICS EXPRESS {\bf 1},  272  (1997).

\bibitem{kunz93}
R. Kunz and R. Berry, Phys.Rev.Lett. {\bf 71},  3987  (1993).

\bibitem{wales94}
D. Wales and R. Berry, Phys.Rev.Lett. {\bf 73},  2875  (1994).
\end{thebibliography}
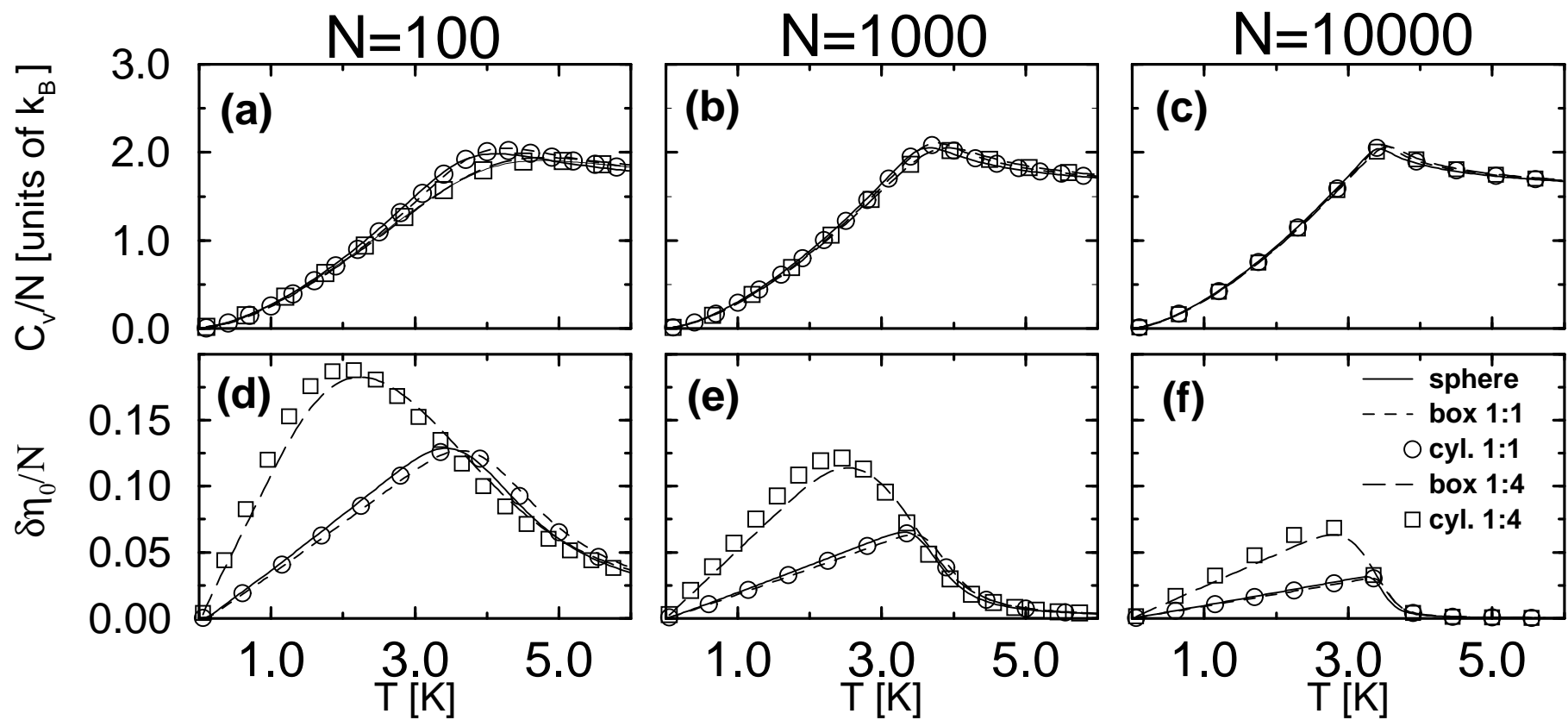
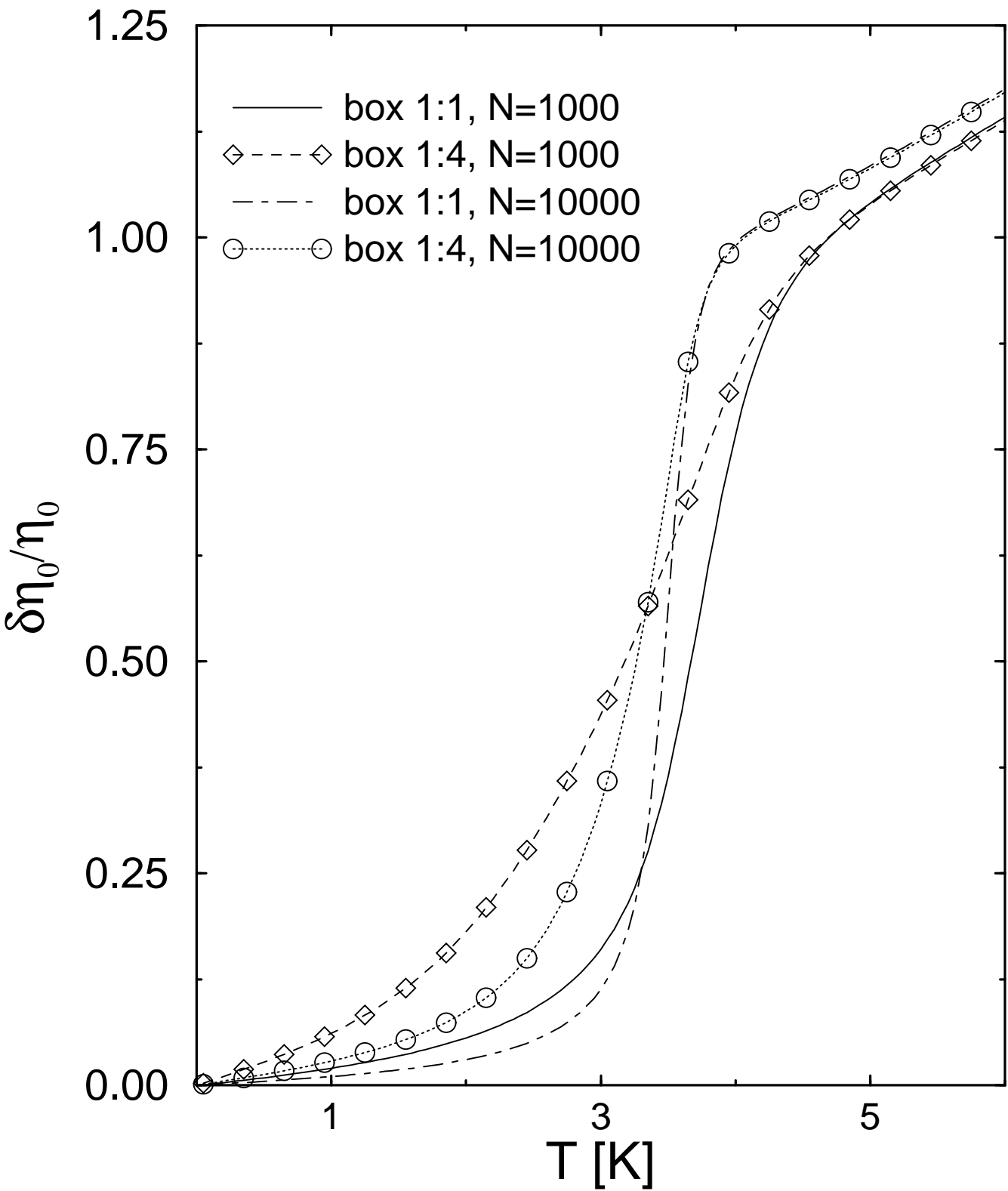
\end{document}